\documentclass[12pt,preprint]{aastex}
\begin{document}
\title{THE DISK-JET CONNECTION IN MICROQUASARS AND AGN} \author{Mario
Livio,\altaffilmark{1} J.~E.\ Pringle,\altaffilmark{1,2} and A.~R.\
King\/\altaffilmark{3}}

\altaffiltext{1}{Space Telescope Science Institute, 3700 San Martin
Drive, Baltimore, MD 21218, USA} 
\altaffiltext{2}{Institute of
Astronomy, Cambridge University, Madingley Road, Cambridge CB3 0HA, UK} 
\altaffiltext{3}{Theoretical Astrophysics Group, University of Leicester,
Leicester LE1 7RH, UK}

\begin{abstract}
We propose a new model for the disk-jet connection in black hole x-ray 
transients and active galactic nuclei. In our model, the inner part of 
the accretion disk around the central black hole switches between two 
states. In one state, the accretion energy is dissipated locally (some 
within the disk, some within a corona) to produce the observed disk 
luminosity. In the second state, the accretion energy is deposited 
into the bulk flow of a relativistic jet. We associate the transition 
between the two states with the generation of a global, poloidal 
magnetic field. We show that this model can explain the observed 
behavior of black hole accretors.
\end{abstract}

\section{Introduction}

Collimated jets are associated with many astrophysical objects,
ranging from active galactic nuclei (AGNs) to young stellar objects
(YSOs). While the precise mechanisms for acceleration and collimation
are still unknown, there is a growing consensus that in most, if not
all cases, jet formation involves an accretion disk threaded by a
large-scale magnetic field. In the case of YSOs, there is direct
observational evidence (in the form of high-resolution images from
\textit{HST}) linking jets to the centers of accretion disks (e.g.\
Burrows et~al.\ 1996). In the case of black hole X-ray transients
(``microquasars'') the evidence is less direct, but is based on
multi-wavelength observations (e.g.\ Mirabel \& Rodriguez 1994).

In the present Letter, we propose a new interpretation of 
microquasar and AGN observations, in the context of a general model that
connects disks and jets through the structure of the poloidal magnetic
field.

\section{Proposed Model}

Belloni et~al.\ (2000) have undertaken a thorough analysis of the X-ray
behavior of GRS~1915+105 as seen by RXTE. They categorize the behavior
by separating it into 12 different classes, and note that it would
have been possible to even split these classes into many more. However, the
main conclusion of the analysis is that the variability can be viewed
as a set of transitions between three basic states. These states (denoted 
by A, B and~C) are separated into three regions of the X-ray color-color
diagram. State~C represents a low luminosity and hard spectral
index. This state corresponds to the low luminosity parts of the
strongly variable behavior (the dips and quiescent phases), and can also
be associated with about half of the time when the source is observed to be in a
constant low state. The spectrum in State~C can be modeled as being 
predominantly a power law with a low energy cut off due to
absorption. There is also a thermalized contribution that can be
interpreted as coming from an accretion disk, but, if so, one has to
assume that the disk emission is truncated at around ten or twenty
inner disk radii (Belloni et~al.\ 1997a). States~A and~B show softer
spectra and can be modeled as corresponding to thermalized emission
from an accretion disk with an inner disk radius of around 20~km. 
State~B is more luminous and hotter than State~A, and can be
interpreted as having a higher accretion rate.

Belloni et~al.\ (1997a,b; 2000) interpret the transition between State~C
and States~A/B as being caused by the disappearance and reappearance
of the inner accretion disk due to a disk instability mechanism of the
kind thought to occur in dwarf novae and other cataclysmic variables.
If such an explanation is to hold, then the instability needs to be of
the ``inside-out'' kind, in which a cooling front starts at the inner disk
edge, progresses out to a radius of some tens of inner radii and then
moves inwards again (Smak 1984). Thus most of the disk remains in the
high state (on the upper branch of the putative $S$-curve in the effective 
temperature-surface density plane) and just the
inner region jumps between the upper and lower branches in the
standard limit cycle fashion. Consequently, during the low/hard state~(C) the
inner part of the accretion disk is either absent or has a much
reduced viscosity, rendering it hard to detect. For the semi-regular
behavior reported in the earlier papers of Belloni et~al.\ (1997a,b)
this appeared to be a reasonable hypothesis, and the time-scales of the
behavior were found to be approximately right.

However, the wide variety in the nature of the variability of the
source reported more recently (Belloni et~al.\ 2000) renders this
explanation more problematic. In particular, for the source to be in
State~C for an extended period of time it would be necessary for the
transition front between the two stable parts of the $S$-curve to be
stationary, which can be shown not to be possible (Smak 1984;
Papaloizou \& Pringle 1985). Instead, we propose here that during
State~C, the inner accretion disk is still present, and is still
accreting at essentially the same rate as before, but, rather, the
energy liberated in the accretion is converted efficiently into
magnetic energy and is emitted in the form of a magnetically dominated
outflow or jet. 

In all models for the transport of angular momentum in
accretion disks (i.e.\ for the viscosity) which involve magnetic fields,
the first step in the process is to convert shear energy in the disk
into magnetic energy, by using a poloidal magnetic flux to
generate a toroidal component (Shakura \& Sunyaev 1973; Eardley \&
Lightman 1975; Galeev, Rosner, \& Vaiana 1979; Pudritz, 1981; Tout
\& Pringle 1992). What then happens to that magnetic energy is still
rather uncertain. Some of the toroidal flux must be used as a
source to generate a poloidal component and so to close the generation
loop for the dynamo, generally assumed to occur through the
magneto-rotational instability (Balbus \& Hawley 1991). But most
of the energy, which comes ultimately from the gravitational potential
energy of the accreting material, must be lost from the disk. There are
three basic possibilities, which are not mutually exclusive. The
first, and the one usually assumed to occur, is that the magnetic
energy is dissipated within the disk. This appears to
happen in the numerical dynamo simulations (Balbus, Hawley, \& Stone
1996) and in this case we have a standard accretion disk that
emits thermalized radiation from its surface (Shakura \& Sunyaev
1973). In this case we expect the disk to be radiating at essentially
all radii.  The second possibility, discussed initially in a number of
contexts (e.g.\ Lynden-Bell 1969; M\'esz\'aros, Meyer, \& Pringle 1977; 
Ionson \& Kuperus 1984), is that the energy might be dissipated locally, 
but instead of within the disk, in the low density regions above and 
below the disk in the form of a chromosphere or corona. Whereas the 
thermalized emission from the disk is radiated at or around the local 
black body temperature, such a corona can reach temperatures comparable 
to the local virial temperature, and can radiate either thermally or by
Compton scattering of low energy photons. Such a model is now regarded
as the standard model for emission from AGN disks (e.g.\ Liang 1979; 
Haardt \& Maraschi 1991) and for the low/hard state of some of the X-ray 
binaries (for example Cyg~X-1; Done et~al.\ 1992). In this case, however, 
unless the radiation process is collimated in some way (for example by 
Compton scattering of soft photons in an outwardly moving bulk flow; 
Malzac, Beloborodov, \& Poutanen 2001) then we still expect to see at 
least some emission from the inner disk regions, for example 
Fe~K$\alpha$ emission caused by irradiation of the disk by
coronal heating. The third possibility is that most of the energy
is not dissipated locally, but is released in the form of
a bulk flow (as a mixture of kinetic energy and Poynting flux) either
as a wind or as a well-collimated jet (Blandford \& Payne 1982). In
this case, it is conceivable that the inner disk radiates at a
severely reduced level, if at all, with essentially all of the
accretion energy being lost in a wind or jet.  We should emphasize
that it is not all the \emph{mass} that is ejected into the jet
(although some is), but most of the accretion \emph{energy}.

\section{The Two States}

The idea that a large fraction of the accretion energy might be
emitted from the disk in the form of an outflow or jet is not
new. Blandford \& Payne (1982) give an idealized calculation in which,
in principle, all of the accretion energy can be emitted as a
collimated outflow (see also Ogilvie \& Livio 2001). Considerations of 
the models for the dynamo process within the disk have led some authors 
to suggest that the conversion of a substantial part of the accretion 
energy to a magnetically dominated outflow might be a standard 
possibility (Kato \& Horiuchi 1986; Tout \& Pringle 1992). There is also
already some circumstantial evidence that the inner regions of some 
accretion disks in close binary systems are under-luminous. In 
particular, a comparison of the temperature profiles (Rutten, van 
Paradijs, \& Tinbergen 1992), and of the mean ultraviolet ($\lambda
\lambda 120$--200~nm) spectra (la~Dous 1991) of the nova-like variables 
(which are thought to be steady standard accretion disks) with those of 
dwarf novae in outburst (quasi-steady disks of comparable brightness),
suggests that the inner parts of the nova-like variable disks are
under-luminous. HST eclipse observations of the nova-like variable
UX~UMa (Knigge et~al.\ 1998) confirm this possibility. These
observations, when taken together, give rise to the idea that the
inner parts of such accretion disks can exist in (at least) two
states. Most of the magnetic field generated by the dynamo process is
small-scale, and tangled within the disk. If the field remains
small-scale then most of the dissipation takes place within the disk
and the disk is radiative. But if a larger-scale field is generated,
perhaps by an inverse cascade above and below the disk (Tout \&
Pringle 1996) then the possibility arises of the more globally
structured, poloidal field being able to drive an energetically
significant outflow. It is important to note that while current MHD 
simulations are generally incapable of following in detail the process 
of large-scale poloidal field generation, the most recent numerical 
results (Turner, Bodenheimer \& Rozyczka 1999; Kudoh, Matsumoto \& 
Shibata 2002) give some credence to the jet production 
process we propose. The concept that an accretion disk in 
a nova-like variable can switch as time proceeds from being initially 
a standard radiative disk to being one in which it releases less of its 
energy thermally finds some support in the observations of the anti-dwarf
nova TT~Ari before and after a short minimum (Tout, Pringle, \& la~Dous
1993). Recent computations on driving outflows in the nova-like
variables IX~Vel and V3885~Sgr also indicate the need for magnetic
driving (Hartley et~al.\ 2002).

\section{Application to GRS 1915+105: The Jet/Disk Connection}

By analogy with the above considerations, we suggest that in
GRS~1915+105 States~A/B correspond to a state of the disk in which a
dynamo-generated field is mainly small-scale with no global poloidal
component. The accretion energy is liberated locally, some within the
body of the disk to generate the thermal component of the spectrum,
and some in a corona to produce the power-law component. We propose,
however, that in State~C, where the inner parts of the accretion disk
appear to vanish, most of the accretion energy is not dissipated
locally, but is emitted in the form of a relativistic jet.

If this is the case, then it should be possible to identify
consequences of the jet emission. Fender (2001) argues that in
addition to the `micro-quasar' GRS~1915+105, four persistent (Cygnus~X$-$1, 
GX~339$-$4, GRS~1758$-$258 and 1E~1740.7$-$2942) and three transient
(GS~2023+38, GRO~J0422+32 and GS~1354$-$64) candidate black hole X-ray
binary systems display persistent flat or inverted radio spectra
occurring during the extended periods in the low/hard state which
corresponds to State~C of Belloni et~al.\ (2000). He suggests that this
spectral component can be identified as synchrotron emission from a
conical, partially absorbed jet (in part, by analogy with similar
behavior in AGN; see, for example, Blandford \& K\"onigl 1979), and
concludes that, if so, the jet power must be a substantial fraction of
the accretion luminosity. 

It might be objected that, in contrast to the AGN case, although some
of the X-ray binaries show direct evidence of jet-like structures
(typically on a scale of parsecs), most do not. However, for a jet to
be visible it is necessary that it radiates, and for it to radiate it
must interact, either internally, or with its environment. There are
three basic zones in which such interactions can occur. The first is at
the base of the jet, where the jet production mechanism is unlikely to
be 100 percent efficient, and is likely to produce a measure of
dissipation. This region is typically of a size of the order of a few
Schwarzschild radii for a black hole accretor, if the jet is to be
relativistic, and so too small to be directly observable. It is likely
however that the dissipation in the jet production process is the
cause of the compact self-absorbed synchrotron emission. The second
zone is external material which the jet might run into. This
interaction zone is clearly the cause of the extended radio emission
seen in radio galaxies, but it seems plausible that most X-ray
binaries are not surrounded by a sufficiently dense cocoon of
material. The third interaction zone is internal to the jet itself and
occurs when variability of the jet leads to internal shocks. It seems
likely that this process is the cause of the observed correlation
between X-ray variability and emission.

Mirabel et~al.\ (1998) report on simultaneous X-ray, infra-red and radio
observations of GRS 1919+105 during which the source oscillated
between States~B and~C on a time-scale of about 30~mins (Class~$\beta$
of Belloni et~al.\ 2001). They associate the emission of IR/radio
emitting plasma with the times of recovery from the X-ray dips; that
is, with the times the source changes from a period in State~C to one
in which the behavior is one of rapid oscillation between States~A/B
and~C. Fender et~al.\ (1999) report on high-resolution MERLIN radio
images of multiple ejections from GRS~1919+105. The superluminal
proper motions of the radio blobs can be traced back to a time when
the source had just emerged from a three-week extended phase in the
low/hard X-ray state and was undergoing oscillations between States~A/B 
and State~C. Feroci et~al.\ (1999) report on the detection of a
large isolated radio flare observed with the Ryle Radio Telescope,
which coincided with a period of oscillatory X-ray behavior seen by
BeppoSAX. Similarly, Fender et~al.\ (2002) again report simultaneous
radio (RATAN$-$600 and ACTA) and X-ray (RXTE) observations in which
large radio flares are seen to emerge at the end of week-long plateau
(State~C) X-ray phases, when again the X-ray behavior is one of rapid
oscillation between States~A/B and State~C.

Taken together, these observational data can be interpreted as
indicating the presence of a steady radio jet during the extended
phases in State~C, with abrupt, enhanced jet emission (flares and the
ejection of blobs) being caused by enhanced variability of the jet
emission mechanism as the disk changes in a stuttering fashion from
putting most of its energy into a jet to dissipating most of its
energy locally. In this context, we draw attention to simultaneous
monitoring of the AGN~3C120 at both X-ray (RXTE) and radio (VLBA)
wavelengths reported by Marscher et~al.\ (2002). Superluminal knots are
seen to emerge along the radio jet in this object. By tracing the blobs
back in time to when they emerge from the nucleus, Marscher et~al.\ conclude 
that each time the X-ray flux dips, and the X-ray spectral index becomes 
harder, a superluminal knot emerges at the site of the radio core.

\section{Modeling the Variability}

Our suggested model for the behavior of GRS~1919+105 depends on the
ability of the central regions of the accretion disk to switch between 
two basic states. We speculate briefly in this section as to how and 
why this might come about. We note that a suggestion of a switching 
behavior which results in a variation of the jet speeds has been 
discussed by Meier et~al.\ (1997). We should also point out that 
while GRS 1919+105 may be regarded as somewhat peculiar, when comparing 
with other black hole systems, it is also the source on which the most 
detailed data are available.

We note first that the launching of a jet probably
requires a poloidal magnetic field of magnitude $B_p$ to exist over a
scale of radius $R$, where $R$ is of order the inner disk radius, and
where $B_p$ is much smaller than the field strengths $B_\mathrm{disk}$
associated with the dynamo-driven magnetic disk viscosity (Pringle 1993). This comes
about by estimating the jet luminosity as being due to the local
poloidal energy density moving outwards at the local dynamical
velocity $v_{\phi}$, i.e. 
\begin{equation}
L_\mathrm{jet} \sim B_p^2 R^2 v_\phi~~,
\end{equation} 
and the accretion disk luminosity (or the rate at which accretion
energy is released in the disk) as
\begin{equation}
L_\mathrm{acc} \sim \dot{M}_\mathrm{acc} v_{\phi}^2~~,
\end{equation}
where $\dot M_{\mathrm{acc}}$ is the accretion rate.

Writing the accretion rate as $\dot{M}_\mathrm{acc} \sim \nu \rho_d H$,
where $\nu\sim\alpha c_s H$ is the viscosity, $\rho_d$ the mean disk
density, $c_s$ the disk sound speed, and $H$ the disk thickness
(assumed $\ll R$), and assuming for the dynamo viscosity that $\alpha
\sim B_\mathrm{disk}^2/\rho_d c_s^2$, we find that (Pringle 1993; Tout \&
Pringle 1996)
\begin{equation}
\frac{B_p^2}{B_\mathrm{disk}^2} \sim \frac{L_\mathrm{jet}}{L_\mathrm{acc}} \frac{H}{R}~~.
\end{equation} 
This calculation is confirmed by Merloni \& Fabian (2002).
Thus, if the jet takes the bulk of the accretion luminosity (i.e.\ $L_\mathrm{jet}
\sim L_\mathrm{disk}$), we require $B_p \sim B_\mathrm{disc} (H/R)^{1/2}$.

Numerical simulations of the dynamo process in accretion disks, while
demonstrating the feasibility of the existence of a self-regulating
dynamo operating within an accretion disk, are unable to undertake the
kind of global computations required to simulate the generation of a
global poloidal magnetic field (although see initial steps in this 
direction by Kudoh et~al.\ 2002). Thus, for the time being we must resort
to physical estimation rather than to detailed modeling.  Using a simple
model for an inverse cascade process, involving reconnection of
magnetic loops above and below the disk, Tout \& Pringle (1996)
demonstrated that such a process might in principle lead to
large-scale fields being generated by the dynamo.\footnote{It should
be noted that previous suggestions that such a poloidal field might be
generated by the dragging in of some global ambient field by the
accretion disk are now generally discounted (Lubow, Papaloizou, \&
Pringle 1994a).} The particular scaling derived by Tout \& Pringle
(1996) indicated that $L_\mathrm{jet}/L_\mathrm{acc} \sim H/R$. However, Livio
(1997)  generalized their analysis and found that it is possible to
have $L_\mathrm{jet}\sim L_\mathrm{acc}$ provided that the distribution $n(l)$ of the
lengths of the magnetic loops is of the form $n(l) \propto l^{-3/2}$.
Thus, while these estimates indicate that the suggestions we put
forward in this paper are not physically unreasonable, as mentioned
above, calculations of the details of this process, including the
time-scales on which changes might occur, are beyond current computing
power.

Instead, we suggest a highly simplified picture of how the relevant
time-scales might be estimated. We consider that the dynamo process
operates locally, so that each annulus of the disk, of radial size
$\sim H$, can be considered to operate independently. This seems a
reasonable conclusion to draw from numerical simulations of 
disk dynamos. This radial size also corresponds to the distance a 
magnetic field can diffuse in the disk using a typical time-scale for the dynamo
cycle of $t_\mathrm{dyn} \sim \mathrm{(a~few)}~\Omega^{-1}$, where $\Omega$ is the local
dynamical frequency, and an effective magnetic diffusivity comparable
to the local effective viscosity (i.e.\ a local effective Prandtl number
of order unity). If we assume that the poloidal field generated by the
dynamo is typically of magnitude $\sim B_\mathrm{disk}$, but of random sign,
spread over of order $m \sim R/H$ azimuthally distributed patches,
then the distribution of net poloidal field from each annulus might be
expected to have a mean value of zero, and a standard deviation of
$\sim B_\mathrm{disk}/m^{1/2} \sim B_\mathrm{disk} (H/R)^{1/2}$. If this magnitude of
poloidal field were reproduced over a range in radius of order $\sim R$,
then, as discussed above, this could give rise to a jet
with $L_\mathrm{jet} \sim L_\mathrm{acc}$. In order to achieve this, we require that $N \sim
R/H$ neighboring annuli all have a locally generated net poloidal
contribution corresponding to of-order-one standard deviation in the
same direction. Since the local dynamos vary on time-scales of $t_\mathrm{dyn}$,
the time-scale on which this is likely to occur is comparable
to tossing $N \sim R/H$ coins once every $t_\mathrm{dyn}$, and asking how
often all the coins come down either all heads or all tails. Thus we
estimate the time-scale for establishing a change in the scale of the 
magnetic field in the disk to be of order
\begin{equation}
t_\mathrm{jet} \sim t_\mathrm{dyn} 2^{R/H}~~.
\end{equation}

If we take $t_\mathrm{dyn} \sim 0.01$~s, corresponding to a couple of 
orbital periods at three Schwarzschild radii around a 10~M$_{\odot}$
black hole, and we take $H/R \sim 0.1$ (Belloni et~al.\ 1997a), then
we obtain $t_\mathrm{jet} \sim 10$~s, which corresponds to the
typical transition time-scale between States~A/B and State~C (Belloni
et~al.\ 1997a,b).

What determines the length of time the source spends in the
various states (for example, how long the source remains in the
low/hard states before switching) is much harder to estimate without a
physical model. The observational data clearly require that both the
high and the low states are reasonably stable for long periods of time,
and so must correspond to equilibria of the system. Inspection of the
various modes of variability categorized by Belloni et~al.\ (2000)
indicates that State~B, that is the state in which there is no jet and
in which all the dissipation occurs locally, is not very stable. All
the light curves displayed in Belloni et~al.\ (2000) show that whenever
the source is predominantly in State~B it switches to and from State~C
in a sometimes erratic and sometimes quasi-periodic fashion. In terms
of our model, this implies that there is always a tendency for the disk
to produce a jet. This tendency appear to be a common one among
systems which contain accretion disks (Livio 2000). On the other
hand, the system spends more than half of the time in State~C and is
able to remain in that state for prolonged periods of time. Thus State~C 
must be a relatively stable equilibrium. For this to
occur we require two conditions. First, once the
dynamo process is able to establish a global poloidal field over a
significant region of the disk, it is able to maintain that field
for a period of time. Numerical simulations of the dynamo process do
show significantly different behavior depending on whether or not a
weak external poloidal field is applied, so there is at least the hope
that this condition can be met. Demonstration of the effect will
require global numerical simulations. Second, the poloidal field must
be established in the center of the disk. This is a restatement of the
effect, noted previously elsewhere (Pringle, 1993; Livio 1997), that the
jet launching process from accretion disks cannot be scale-free and
that jets are in general launched from the central disk regions.

In this regard we draw attention to two physical effects. The first is
that a poloidal magnetic field can diffuse radially through a
turbulent accretion disk on a time-scale $H/R$ times shorter than the
radial flow velocity due to the magnetic viscosity (Lubow et~al.\ 
1994a). This is the reason (mentioned above) why it is 
generally not possible to drag an external ambient field inwards in an
accretion disk. The second effect works in the opposite way: if
the poloidal field gives rise to a jet/wind which carries away angular
momentum at a rate comparable to or greater than that carried by the
magnetic viscosity, then the inflow speed in the disk is significantly
enhanced. Thus, for example, if in the central disk regions where the
jet is launched the inflow speed caused by angular momentum loss in
the outflow is increased over the speed due to the magnetic viscosity
by a factor of $\sim R/H$, then it becomes possible to trap the poloidal
flux generated by the dynamo within those regions. This would enable
to jet to be launched predominantly from the center of the
disk. However, this same effect, also implies (Lovelace, Romanova, \& Newman 1994; Lubow
et~al.\ 1994b) that if a poloidal field is established over a
significant region in the outer part of the disk, to the extent that the
outflow generated by it causes significant angular momentum loss, then
an instability develops whereby those parts of the disk fall rapidly
inwards. This probably implies that it is not possible to establish a
steady jet from the outer parts of the disk. However, there is no
reason why the dynamo process should not be able to establish a
significant global poloidal field in the outer parts of the
disk. This would occur on a longer time-scale than that in the inner regions,
because the dynamical time-scales are longer at larger radii, and might
be the cause of some of the longer term erratic variability. For
example, if an outer disk region generates and advects into the
center a poloidal field of opposite polarity (to that already in the
center and generating a jet), then this might be a mechanism for a
sudden switch-off of the jet and a sudden switch from State~C to State~B.

The switch in magnetic viscosity from radiative dissipation to driving
outflow discussed here offers a possible explanation for the very long
recurrence times of soft X-ray transients (SXTs). SXT outbursts
themselves are convincingly explained as thermal-viscous disk
instabilities, greatly prolonged by irradiation by the central
X-rays, which trap the disk in the hot, high-viscosity state until
most of its mass has been accreted (King \& Ritter 1998; in the case 
of GRS 1915+105 the limit cycle may operate between the standard branch 
and the slim-disk branches, e.g.\ Honma, Matsumoto \& Kato 1991). But it is
currently unclear why the outbursts recur only at intervals of 
$t_\textrm{rec} \sim 1$--50~yr or more. In a standard picture where viscosity
produces dissipation, the central disk regions would refill on a short
viscous timescale, and would thus reach the critical surface density
for triggering an outburst on much shorter timescales $t_\textrm{rec}
\sim$ weeks to months, like those of dwarf novae with comparable disk
sizes (and thus similar binary periods). However, if instead, viscosity
drives significant outflow, the central disk regions may remain much
more depleted than in a dwarf nova, giving much longer recurrence
times. A direct consequence of this idea is that quiescent SXTs may
show significant radio emission. It appears that this is indeed the
case for V404~Cygni (observations by R.~Hjellming: M.~Rupen, private
communication). It would evidently be worthwhile checking other
quiescent SXTs for radio emission. Very interestingly, Sokoloski \&
Kenyon (2002) point out that the symbiotic binary CH~Cyg shows
a radio jet in an optical low state, and indeed behavior closely
analogous to that described above for GRS~1915+105. This supports the
view (Livio 1997, 2002; Price, Pringle, \& King, 2002) that
there may be a common, accretion-powered formation mechanism for all
jets.  We also note that recent observations of the radio galaxy IC~4296 
(Pellegrini et~al.\ 2003), and of M~87 (Di~Matteo et~al.\ 2003), find 
that a sizable fraction of the accretion energy in these sources goes 
into powering the jet, in support of the scenario presented in this paper.

\end{document}